\documentclass[twocolumn]{aastex631}
\usepackage{enumerate}

\usepackage{tikz}
\usetikzlibrary{shapes.geometric, arrows}

\tikzstyle{startstop} = [
    regular polygon, 
    regular polygon sides=6, 
    shape border rotate=180, 
    minimum width=3cm, 
    minimum height=1cm, 
    text centered, 
    draw=black, 
    fill=white, 
    text width=2.5cm
]
\tikzstyle{process} = [
    regular polygon, 
    regular polygon sides=6, 
    shape border rotate=180, 
    minimum width=3cm, 
    minimum height=1cm, 
    text centered, 
    draw=black, 
    fill=white, 
    text width=2.5cm
]
\tikzstyle{arrow} = [thick,->,>=stealth]

\begin{document}

\title{Illuminating Youth: Decades of Mid-Infrared Variability and Color Evolution of Young Stellar Objects}

\correspondingauthor{Neha S.}
\email{pathakneha.sharma@gmail.com}


\author{Neha S.}
\affiliation{Aryabhatta Research Institute of Observational Sciences, Nainital\textendash263001, Uttarakhand, India}
\author{Saurabh}
\affiliation{Aryabhatta Research Institute of Observational Sciences, Nainital\textendash263001, Uttarakhand, India}

\begin{abstract}
The variability of Young Stellar Objects (YSOs) is a crucial tool for understanding the mechanisms driving flux changes. In this study, we present an infrared variability analysis of a large sample of over 20,000 candidate YSOs, using data from the ALLWISE and NEOWISE surveys, which span around a decade with a 6-month cadence. We applied Lomb-Scargle Periodogram (LSP) analysis and linear fitting to the light curves, classifying them into distinct categories: {\it Secular} ({\it Linear},  {\it Curved}, and  {\it Periodic}) and  {\it Stochastic} ({\it Burst},  {\it Drop}, and  {\it Irregular}). Our findings show that 5,467 (26.2$\pm$0.3\%) of the sources exhibit variability, with most (19.7$\pm$0.3\%) showing {\it Irregular} variations, followed by {\it Curved} and {\it Periodic} variations. In addition, 235 sources of {\it Bursts} and 122 {\it Drop} sources were identified. Variability is more pronounced in Class I sources with a higher fraction of variables (36.3$\pm$0.6\%) compared to Class II (22.1$\pm$0.4\%) and Class III (22.5$\pm$1.0\%) sources. The color (W1 $-$ W2) versus magnitude analysis (W2) using linear fitting shows that the trend ``redder-when-brighter" (RWB) is more prevalent (85.4$\pm$0.5\%) among YSOs. In contrast, the trend ``bluer-when-brighter" (BWB) is more common in younger sources compared to more evolved ones, having a BWB fraction of 29.0$\pm$1.1\% for Class I to 4.0$\pm$0.9\% for Class III.

\end{abstract}


\section{Introduction} \label{sec:intro}
Brightness variability is a prevalent phenomenon observed in Young Stellar Objects (YSOs), offering valuable insights into their dynamic nature and underlying physical mechanisms. The observed changes in YSO brightness stem from many mechanisms impacting the star, disk, and surrounding envelope. These mechanisms include rotational modulation of stellar surface features, unstable accretion processes, variations in extinction due to surrounding material, disk instabilities, and the presence of eclipsing binary systems \citep{1992Attridge,1996Choi,2013Romanova,2014Stauffer,2013Bouvier,2023Bino,2004Stassun,2008Cargile}.

Studies spanning optical, near-infrared, mid-infrared, and sub-millimeter wavelengths have shed light on the intricate processes shaping YSO behavior \citep{1994Herbst,2001Carpenter,2021LeeYH}. At longer wavelengths, such as the mid-infrared, observations provide a unique window into the properties of deeply embedded protostars within their nascent envelopes. These observations are particularly significant for studying the youngest protostars, often obscured at shorter wavelengths \citep{2007A&A...470..211K, 2015ApJ...800L...5S, 2018ApJ...854..170H, 2018ApJ...863L..12L}. The variability observed at millimeter wavelengths, attributed to temporal changes in protostellar luminosity, presents valuable insights into the accretion processes driving stellar growth \citep{2013ApJ...765..133J, 2019MNRAS.487.5106M, 2020ApJ...895...27B, Contreras20}.

The Young Stellar Object VARiability (YSOVAR) program, enabled by the Spitzer Space Telescope, has unveiled the complexity of YSO variability, highlighting the diversity of physical mechanisms driving changes in the mid-infrared \citep{morales11, 2014Cody, apjad14f8bib31, 2018Wolk}. Long-term mid-infrared variability has recently been conducted using NEOWISE. These studies revealed the primary mechanisms driving variability, including changes in extinction along the line of sight and variable accretion \citep{park21}. The short timescale variability (up to $\sim$40 days) has been extensively studied through programs like YSOVAR, highlighting rotational modulation, variable extinction, and stochastic bursts as prominent factors contributing to YSO variability \citep{morales11, 2014Cody, apjad14f8bib31, 2018Wolk, 2024Lee}. In comparing short and long-term variability, \cite{2024Lee} found that the long-term variability amplitude is, on average, a factor 3 larger than the short-term variability, and variability amplitude increases as the timescales increase.

This work aims to shed more light on the YSO variability with a much larger sample using archival data from ALLWISE and NEOWISE surveys spanning about a decade of observations. The primary objectives are to check the pattern exhibited by YSOs in different evolutionary stages and understand the origin of the variation in the light curve, separating the sources into different types: {\it Secular} ({\it Linear}, {\it Curved}, and {\it Periodic}) and {\it Stochastics} ({\it Burst}, {\it Drop}, and {\it Irregular}) sources. Furthermore, we investigate the variation in color with the change in magnitude of sources at different stages. We made the entire catalog available publicly to facilitate further study of YSOs. The paper is structured as follows. In section \ref{sec:data}, we present our sample and how the data is collected. In section \ref{sec:results}, we perform several analyses to find out variable sources and then study their lightcurve behavior; in section \ref{sec:discussion}, we discuss our findings, and finally, we conclude our work in section \ref{sec: conclusion}.

\section{Sample and Data} \label{sec:data}
We used the Spitzer/IRAC Candidate YSOs (SPICY) catalog of \cite{2021ApJS..254...33K}, which includes 117,446 YSO candidates selected from the Spitzer/IRAC survey. The sources are between $l\sim255$ degrees and 110 degrees in the Galactic midplane. The catalog uses a statistical learning method that uses Spitzer's spatial resolution and sensitivity in the mid-infrared range and separates YSOs using Infrared Array Camera (IRAC) of Spitzer four-band infrared color excess selection criteria. They obtained 117,446 candidate YSOs and 180,997 probable contaminants among their sources with infrared color excess. The candidate YSOs exhibit a highly structured spatial distribution on the galactic mid-plane, revealing cluster-like and filament-like patterns, and are located at distances larger than 1 kpc. The YSO candidates exhibit a smooth distribution in the IRAC color space, with  3.6$\micron$ to 4.5 $\micron$ color peaks at approximately 0.5 mag and 4.5$\micron$ to 5.8$\micron$ bands color near 0.4 mag. \citet{2021ApJS..254...33K} also noted that the majority of the YSOs are young pre-main sequence stars having ages around 1-10 Myr.

\begin{figure}
    \centering
    \includegraphics[width=9cm, height=10cm]{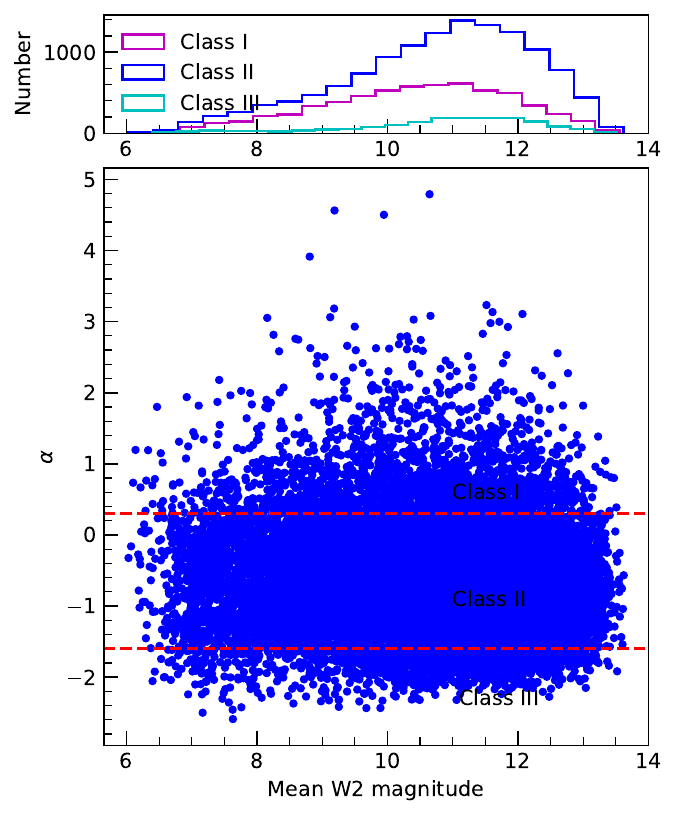}
    \caption{Spectral index is plotted against mean W2 magnitude of our sources. The horizontal lines separate the YSOs into different classes. The W2 (mean) band magnitude distribution is shown at the top.}
    \label{fig:mag}
\end{figure}

The catalog also provides YSO type (evolutionary stages) from the multi-band (4.5 to 24 micron) SED slope and classified the sources in Class I (15,943), flat spectrum (23,810), Class II (59,949), and Class III (5352), apart from that the sources with missing photometry are classified as uncertain type. In Figure \ref{fig:mag}, we plotted the mean magnitude of our sample against the spectral index \citep[$\alpha$ taken from SPICY catalog of][]{2021ApJS..254...33K} separately for different YSO Classes. Most of our sources have a magnitude range of 10-12 mag in the W2 band.

Since we aim to study the infrared variability, we cross-matched all these sources in the WISE catalog with a search radius of 2 arcsec. The coordinate search of the sources with the WISE database resulted in 42501 sources for further analysis. This significant reduction in the sample is mainly because the parent SPICY catalog is made of infrared photometric measurements from the four-channel of IRAC as it offers much higher spatial resolution ($\sim$1.7 arcsec while WISE has a resolution of ~6 arcsec) of any Mid-IR imaging camera in the wavelength range of 3-9 micron with wide-field imaging capability. Furthermore, the sensitivity of IRAC is higher in the galactic plane as WISE is limited by detector saturation and source confusion. 

The WISE survey \citep{2010AJ....140.1868W} includes data from the ALLWISE survey, which consists of ALLWISE Cryogenic survey (Jan 2010$-$Aug 2010), WISE 3-band Survey (Aug 2010$-$Sep 2010) and NEOWISE Post-Cryogenic Mission\footnote{\url{https://irsa.ipac.caltech.edu/Missions/wise.html}} (Sep 2010$-$Feb 2011) \citep{2011ApJ...731...53M}. Then, after a gap, the NEOWISE Reactivation survey started in Dec 2013 and continued observations until mid-2024 \citep{2014ApJ...792...30M}. Apart from the ALLWISE data (2009$-$2010), we have used NEOWISE data from the NEOWISE 2023 data release that contains all NEOWISE observations between Dec 12, 2013 to December 13, 2022, UTC in 3.4 (W1) and 4.6 (W2) micron bands. Therefore, this study includes about ten years of WISE observations. 

WISE continuously stares at a portion of the sky every six months for 1-2 days. Therefore, long-term variability with a cadence of 6 months and short-term variability for 1-2 days with a cadence of 1.5 hours is possible. Since we combine ALLWISE and NEOWISE data, we can study the long-term variation of about a decade. Several factors can contaminate WISE photometry; therefore, we have selected good-quality exposure adopting several criteria\footnote{\url{https://wise2.ipac.caltech.edu/docs/release/allwise/expsup/sec2\_1a.html\#nb}}: 1)  saa\_sep $> 0.0$, `moon\_masked' = `0000', `qi\_fact' $>0$, nb' $<$ 3, `na' $=$ 0, `cc\_flags' = `0000' and uncertainty in W1 and W2 mag per exposure is lower than 0.2 mag. The above quality cut further reduces the sample to 36353 YSOs.   

\begin{figure}[h] 
    \centering
    \vspace{10pt} 
\begin{tikzpicture}[
    node distance=2cm, 
    every node/.style={align=center, font=\small}, 
    process/.style={rectangle, rounded corners, draw=black, fill=blue!20, text width=7cm, minimum height=1cm, text centered},
    startstop/.style={rectangle, rounded corners, draw=black, fill=red!20, text width=7cm, minimum height=1cm, text centered},
    arrow/.style={thick,->,>=stealth}
]
\node (start) [startstop] {\textbf{Initial Sample:} \\ 117,446 sources};
\node (step1) [process, below of=start] {\textbf{WISE Database Search:} \\ 42,501 YSOs identified};
\node (step2) [process, below of=step1] {\textbf{Quality Cuts Applied:} \\ Filtered to 36,353 high-quality light curves};
\node (step3) [process, below of=step2] {\textbf{Light Curve Selection:} \\ Uncertainty in the individual epochs lower than 0.2 mag and at least 16 points over 8 years \\ $\rightarrow$ 20,893 YSOs retained};
\draw [arrow] (start) -- (step1);
\draw [arrow] (step1) -- (step2);
\draw [arrow] (step2) -- (step3);
\end{tikzpicture}
\vspace{5pt} 
    \caption{Flowchart illustrating the filtering process of YSOs from the initial sample to the final dataset.}
    \label{fig:ysos_flowchart} 
\end{figure}
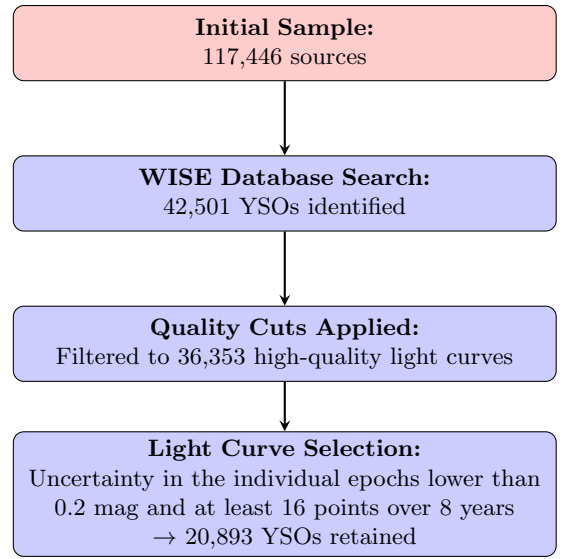

To create a long-term lightcurve, we first removed the outliers by performing a 3-$\sigma$ clipping, i.e., 3 times the standard deviation, around the median of the data points within each 6-month window (therefore, one epoch in every six months). Then, we have taken an average of all the data points within the 6-month window to make a long-term lightcurve with a maximum duration of $\sim$10 years containing a minimum of 16 points and a maximum of 21 points. The majority of the objects have more than 18 points. We finally considered uncertainty in the individual epochs lower than 0.2 mag and a minimum of 16 epochs on the averaged light curve for the variability and color analysis, which resulted in the final sample of 20,893 YSOs. The entire process has been shown in Figure \ref{fig:ysos_flowchart}. Therefore, our final sample size is primarily dictated by the WISE sensitivity, resolution, and sampling.

The uncertainty at each epoch\footnote{Here, epoch refers to the average of all the data points within a 6-month window.} ($\Sigma_j$) in the 6-month averaged light curve is the quadratic addition of the standard deviation of the fluxes within the short-term light curve (6-months) and the uncertainty of individual flux points within it, i.e., 
\begin{equation}\label{eq:sig}
    \Sigma_j^2  = \frac{1}{N-1} \sum_{i=1}^{N} (m_i - \bar{m})^2 + \frac{1}{N^2} \sum_{i=1}^{N} \sigma_i^2
\end{equation}
where $m_i$ and $\sigma_{i}$ are the magnitude and uncertainty of $i$-th exposure/frame, $\bar{m}$ is the average magnitude of the epoch, and N is the total number of single exposures within the 6-month window. The overall uncertainty ($\sigma$) of the final light curve is calculated as the mean of the uncertainty of individual epochs i.e., $\sigma = <\Sigma_j>$ as calculated via equation \ref{eq:sig}. We have used the W2 band for the variability study as YSOs are fainter in the W1 bands. However, we performed color-magnitude analysis using W1 and W2 bands.

\begin{figure}
    \centering
    \includegraphics[width=8.5cm, height=9.5cm]{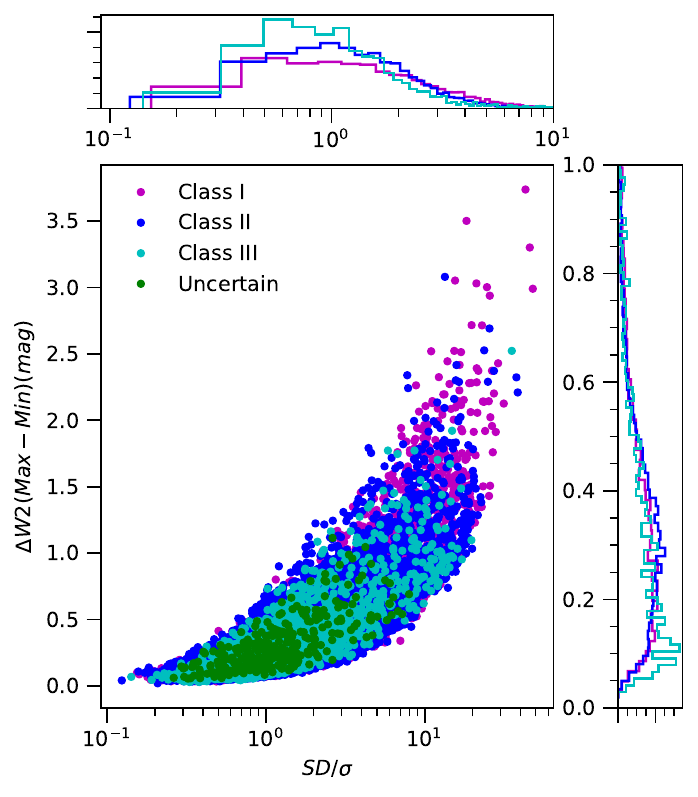}
    \caption{The fractional flux change between bright and faint epochs $\Delta W2$ (y-axis) is plotted against the ratio of the standard deviation of the light curve and the mean flux uncertainty. Different classes of YSOs are plotted. The 1D distribution is shown at the top and right with a limited x-range for clear visibility. The 1D distribution of the `Uncertain' type is not shown due to low number statistics. }
    \label{fig:var}
\end{figure}

\section{Results and analysis} \label{sec:results}
To quantify the variability amplitude, we calculated the fractional flux change between the brightest and faintest epoch for a given lightcurve, i.e., $\Delta W2 =|\mathrm{min}(W2) - \mathrm{max}(W2)|$.  Previous studies on YSO variability use the standard deviation of the light curve and mean measurement error in the flux scale to quantify the variability \citep{2021LeeYH}. Therefore, we first converted the magnitude into flux\footnote{The zero points for converting the magnitude to flux is taken from the WISE manual \url{https://wise2.ipac.caltech.edu/docs/release/allsky/expsup/sec4_4h.html}.} and then calculated the standard deviation and mean error. We followed the recent work by \citet{2024Lee} and classified sources into different variable categories \citep[see also][]{park21}. We applied two criteria to select the variables: I) $\Delta W2$/$\sigma >3$. This condition gives us 16704 variables out of 20893 YSOs studied (i.e., 80\%) and II) SD/$\sigma$ $>3$, which provides us 4803 (23\%) variables, all of which are also satisfying the above first (I) condition. Moreover, we found 930 sources with $\Delta W2 >1$ mag and 68 sources with $\Delta W2>2$ mag in our sample. We plotted all the sources in Figure \ref{fig:var} where sources are separated into different categories: Class I (Combined Flat-Spectrum and Class I; 6159), Class II (12757), Class III (1659), and Uncertain (318). The median variability amplitude (SD/$\sigma$) is 2.0 (Class I), 1.6 (Class II), 1.3 (Class III), and 1.1 (Uncertain). The median $\Delta W2$ is found to be 0.34 mag (Class I), 0.32 mag (Class II), 0.29 mag (Class III), and 0.25 mag (Uncertain). A higher SD/$\sigma$ implies a high $\Delta W2$. However, on average, Class II and III objects show similar variability amplitudes that is lower than Class I objects. 

\begin{figure}
    \centering
    \includegraphics[width=9cm, height=7cm]{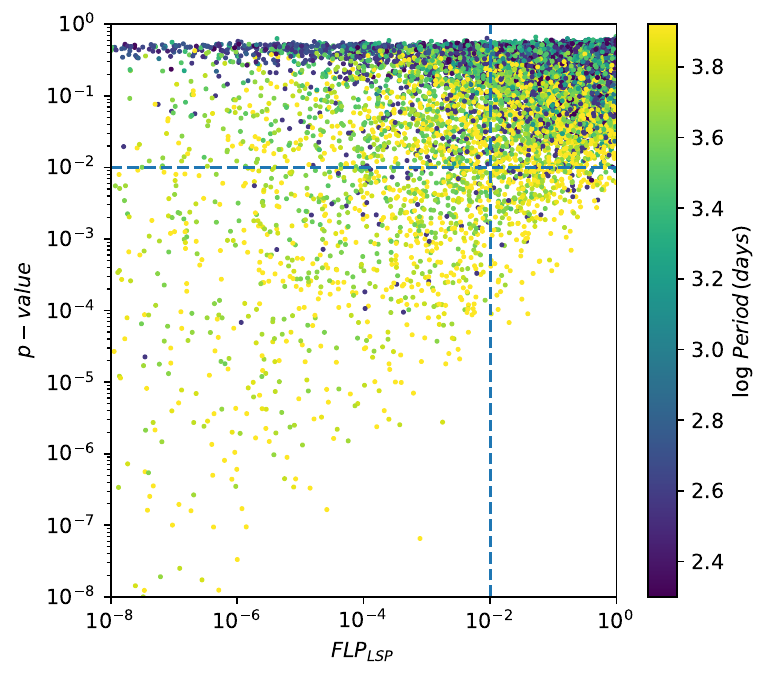}
    \caption{The $p$-value of the null hypothesis is plotted against the False alarm probability of LSP. The vertical and horizontal lines indicate the limit to classify the YSO variability into different types.}
    \label{fig:linear-periodic}
\end{figure}

\begin{table*}
    \centering
    \resizebox{1\textwidth}{!}{ 
    \begin{tabular}{c|c|c|c|c|c|c|c|c} \\ \hline \hline 
    YSO Class  &  No of objects &  {\it Linear} & {\it Curved} & {\it Periodic} & {\it Burst}  & {\it Drop} & {\it Irregular} & {\it Non-variable}  \\ 
    (1) & (2) & (3) & (4) & (5) & (6) & (7) & (8) & (9) \\  \hline
    Total sample     &  20893 & 216 ($1.0\pm0.1$) & 589 ($2.8\pm0.1$) & 190 ($0.9\pm0.1$) & 235 ($1.1\pm0.1$) & 122 ($0.6\pm0.1$) & 4115 ($19.7\pm0.3$) & 15426 ($73.8\pm0.3$) \\ 
    Class FS/I &  6159  & 113 ($1.8\pm0.2$) & 243 ($3.9\pm0.2$) & 61 ($1.0\pm0.1$)  & 80 ($1.3\pm0.1$)   & 29 ($0.5\pm0.1$)  & 1708 ($27.7\pm0.6$) & 3925 ($63.7\pm0.6$) \\
    Class II   &  12757 & 78 ($0.6\pm0.1$)  & 290 ($2.3\pm0.1$) & 106 ($0.8\pm0.1$)  & 118 ($0.9\pm0.1$)  & 69 ($0.5\pm0.1$)  & 2157 ($16.9\pm0.3$) & 9939 ($77.9\pm0.4$) \\
    Class III  &  1659  & 24 ($1.4\pm0.3$)  & 51 ($3.1\pm0.4$)  & 22 ($1.3\pm0.3$)  & 30 ($1.8\pm0.3$)   & 24 ($1.4\pm0.3$)  & 223 ($13.4\pm0.8$)  & 1285 ($77.5\pm1$) \\
    Uncertain  &  318   & 1 ($0.3\pm0.4$)   & 5 ($1.6\pm0.7$)   & 1 ($0.3\pm0.4$)   & 7 ($2.2\pm0.8$)    & 0 ($0.0\pm0.2$)   & 27 ($8.5\pm1.6$)    & 277 ($87.1\pm1.9$) \\ \hline \hline
    \end{tabular}  }
    \caption{Type of variability based on YSOs Classification. The number in the parenthesis represents the fraction of sources to the total number of sources (mentioned in column no. 2 of the respective rows) in percentage. Here, uncertainties in the fraction are determined using an approximate 68\% binomial confidence interval \citep{Wilson01061927} after averaging the asymmetric confidence interval. }
    \label{tab:var_stat} 
\end{table*}

\begin{figure*}
    \centering
    \includegraphics[width=18cm, height=10cm]{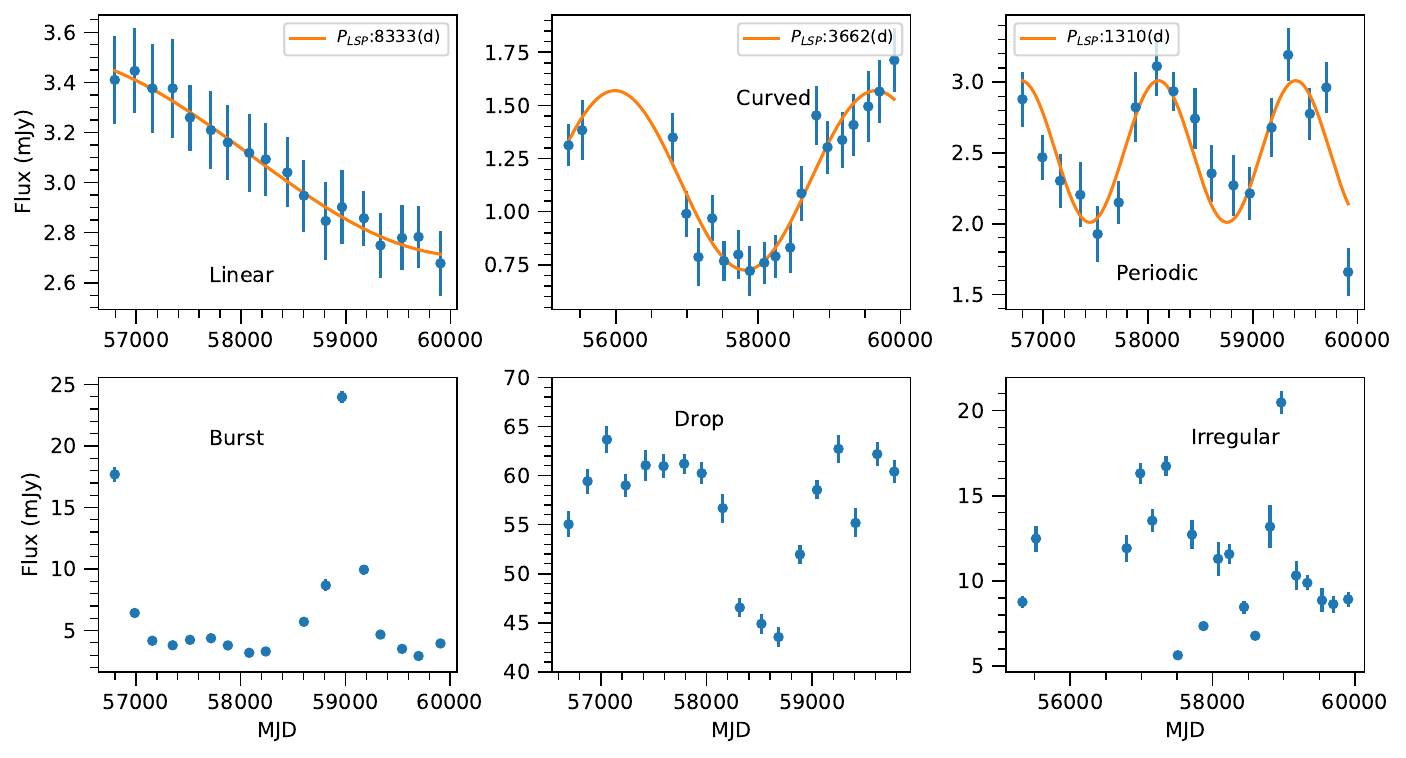}
    \caption{Example light curves of different types of variables: {\it Linear}, {\it Curved}, {\it Perodic}, {\it Burst}, {\it Drop}, and {\it Irregular} from left to right. The LSP fitting and its period are shown in the upper panels. Here, the error bars are the uncertainty at individual epochs in the final light curve as calculated using equation \ref{eq:sig}.}
    \label{fig:var_type}
\end{figure*}

\subsection{Variability classification}
We used the Lomb-Scargle Periodogram \citep[LSP][]{apjad14f8bib19, apjad14f8bib27} method to analyze our light curves. This method is commonly used for identifying periodic signals in unevenly sampled time-series data. We used the ``LombScargle" module from the ``astropy" package of Python. To quantify the uncertainty of a particular LSP peak, we calculated the false alarm probability (FAP\_LSP) using the bootstrap method \citep{efron1993bootstrap} simulating 10,000 re-sampled light curves for each observed light curve. 

Since the typical cadence of our long-term light curve is $\sim$6 months and at least two periods are needed to confirm a periodic variable, we cannot confirm a periodic pattern in a lightcurve having periodicity less than 6 months or more than 4.5 years from the current data sets. Objects with very long periods, twice the maximum duration, i.e., 18 years, can be considered to have a linear pattern. These sources can also be represented by linear fit with increasing/positive or decreasing/negative slope. Therefore, we also performed linear regression and estimated the Pearson correlation coefficient and $p$-value for a hypothesis test whose null hypothesis slope is zero. We have plotted the $p$-value vs. FAP\_LSP in Figure \ref{fig:linear-periodic}.

The standard deviation of a light curve is expected to be large when the variability significantly exceeds the measurement uncertainty. However, short-term variations occurring in only a few epochs may not result in a high standard deviation. Additionally, long-term trends, including gradual or periodic variations, can be present in the light curve while still exhibiting a low standard deviation. Therefore, we adapt the first criteria and study the variability pattern for all the 16704 sources in the light curve. In Table \ref{tab:var_stat}, we provide the number statistics of different variables. The detailed lightcurve statistics for individual objects are provided in Table \ref{tab:var_stat_full_table}. In total, we found 5467 ($\sim 26.2\pm0.3$\%) sources to be variable.

\begin{figure}
    \centering
    \includegraphics[width=9cm, height=9cm]{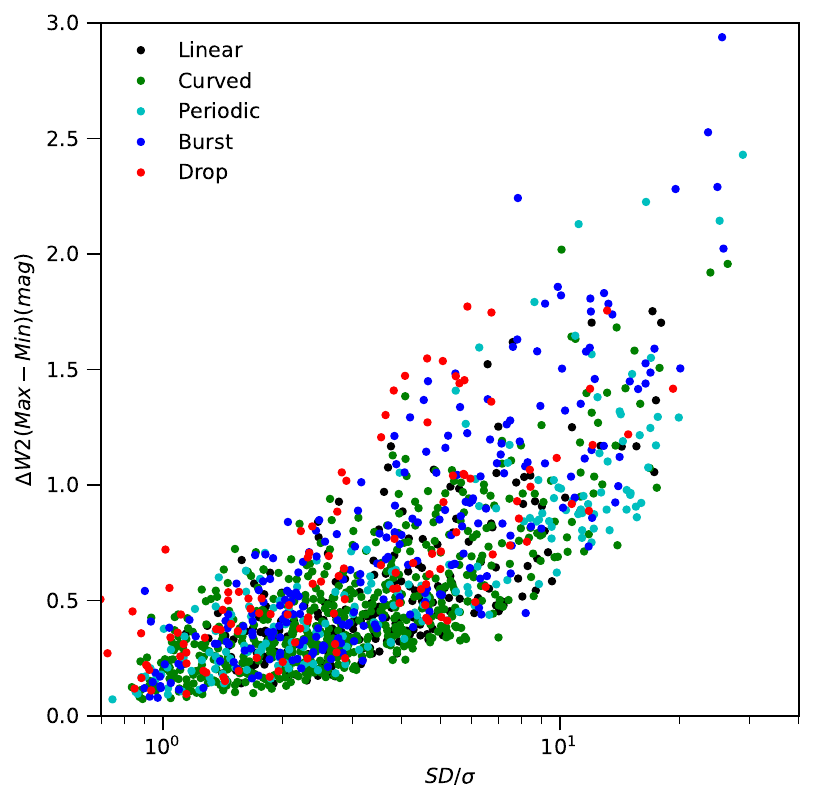}
    \caption{$\Delta W2$ is plotted against SD/$\sigma$ for different types of variables. {\it Irregular} variables are not shown due to their large sample size.}
    \label{fig:var_yso_type}
\end{figure}

In the following subsection, we will discuss the variables classified into different categories based on their light curves' shape and periodicity.

\subsubsection{Linear}
{\it Linear} variables are those with linearly increasing (positive slope) or decreasing (negative slope) trends in their light curve. {\it Linear} was identified using the conditions of False alarm probability of LSP $<0.01$ and having a very large period of 18 years (2 times nine years) and p\_value of the null hypothesis in the linear regression $<0.01$. About 216 ($1\pm0.1$\%) sources in our sample show linear trends. An example of a {\it Linear} light curve is shown in the top-left panel of Figure \ref{fig:var_type}.  

\subsubsection{Curved}
The LSP method enables us to find periodic variables; however, the light curve should cover at least two cycles to confirm their periodicity. Considering the average length of the lightcurve is ~9 years, we considered sources to be {\it Curved} if they have longer periods between 1643 (4.5 years) days to 6570 (18 years) and FAP\_LSP $<0.01$. We found 589 ($2.8\pm0.1$\%) sources in our sample to show curved behaviors in their light curves.  Class I sources show more curved behavior compared to Class II and III sources. The top-middle panel of Figure \ref{fig:var_type} shows an example of a curved light curve.

\subsubsection{Periodic}
Periodic variables are those having FAP\_LSP $<0.01$ and LSP periods between 200 and 1643 days. We found 190 ($0.9\pm0.1$\%) {\it Periodic} variables in our sample. Figure \ref{fig:var_type} shows an example of the periodic light curve with a period of 1338 days. A sinusoidal function with the derived period is overplotted to demonstrate the pattern in the light curve variability. More Class III sources are found to show periodic behavior compared to other classes. An example of a periodic light curve is presented in the top-right panel of Figure \ref{fig:var_type}.   

\subsubsection{Burst}
{\it Burst} variables exhibit intermittent brightness peaks amid stable fluxes across epochs. To discern {\it Bursts}, we set a threshold $\Delta W2$/$\sigma >$ 3 to isolate significant flux increases. We introduce an additional criterion involving $\Delta W2$, median, and minimum magnitudes: a target qualifies as a {\it Burst} if $|$(median(magnitude)–maximum magnitude)$| >$ 0.8 $\times \Delta W2$. We found 235 ($1.1\pm0.1$\%) {\it Burst} sources in our sample. The bottom-left panel of Figure \ref{fig:var_type} represents a {\it Burst}-type light curve.

\subsubsection{Drop}
The {\it Drop} variable is the opposite of the {\it Burst} variable, i.e., sources exhibiting intermittent faintness peaks amid stable fluxes across epochs. Therefore, {\it Drop} variables are identified by $|$(median(magnitude)–maximum magnitude)$| <$ 0.8 $\times \Delta W2$. We found 122 ($0.6\pm0.1$\%) {\it Drop} variables in our sample. The bottom-middle panel of Figure \ref{fig:var_type} shows an example light curve for {\it Drop}-type variability.

\subsubsection{Irregular}

After identifying five different types of variables from {\it Linear} to {\it Drop} from all 16704 YSOs by applying $\Delta W2/\sigma >3$, the number of remaining sources is 15352. The light curves of these remaining targets show random behavior. We finally adopted the general condition for variability, SD/$\sigma >3$, to identify variables with {\it Irregular} light curves. We found 4115 ($19.7\pm0.3$\%) {\it Irregular} variables in our sample. An example light curve for {\it Irregular} is presented in the bottom-right panel of Figure \ref{fig:var_type}. 

Figure \ref{fig:var_yso_type} shows the relationship between $\Delta W2$ and SD/$\sigma$ for different types of variables (except {\it Irregular} due to their large sample size). The {\it Stochastic} sources exhibit a higher median $\Delta W2$ of 0.60 mag, in contrast to the {\it Secular} sources, which have a median of 0.42 mag. Among the {\it Stochastic} sources, {\it Bursts} demonstrate a higher average variability amplitude, with a median value of 0.65 mag. Out of 68 sources with $\Delta W2 > 2$, 62 sources exhibit {\it Stochastic} variability. Note that some of these variables could be of a combined type, i.e., a mixture of a long-term secular trend with a {\it Burst} or {\it Drop} feature \citep[e.g.,  see][]{2021LeeYH}.

\begin{figure}
    \centering
    \includegraphics[width=9cm, height=7cm]{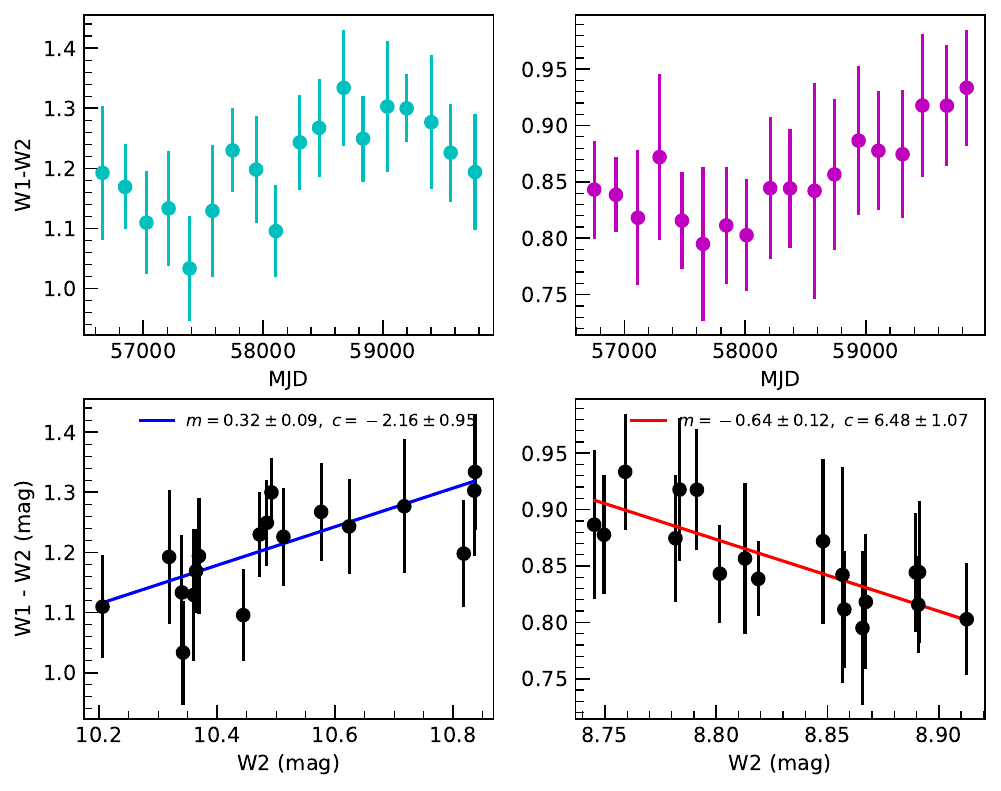}
    \caption{An example of BWB (left) and RWB (right) is shown. The magnitude is shifted in the 1st panel for better visibility.}
    \label{fig:example_color}
\end{figure}

\begin{figure*}
    \centering
    \includegraphics[width=16cm, height=8cm]{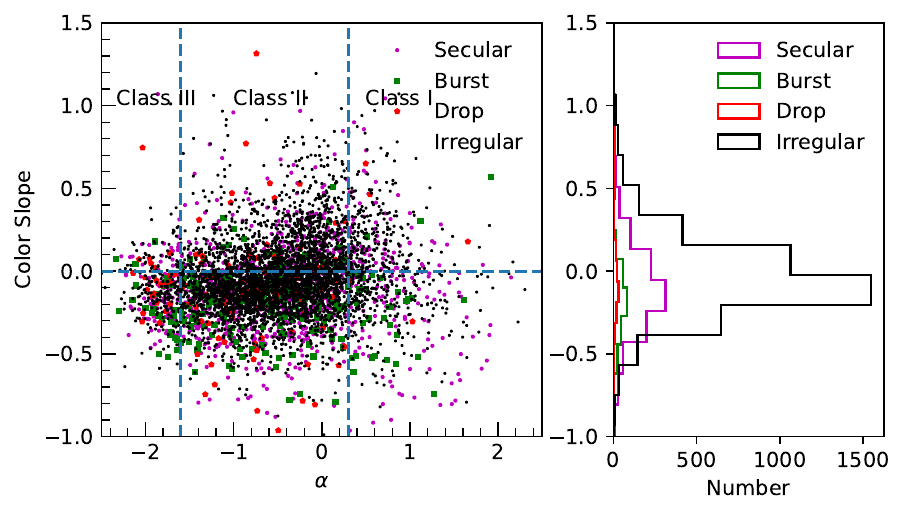}
    \caption{Left: The correlation between slope in color-magnitude diagram vs. spectral index (YSO class). {\it Secular} ({\it Linear}, {\it Curved},  and {\it Periodic}), {\it Burst}, {\it Drop}, and {\it Irregular} sources are also marked. The horizontal line denotes zero colors. Right: The slope of line fit of the color (W1-W2) vs W2 mag diagram. A negative slope indicates a redder-when-brighter (RWB) slope, while a positive slope indicates a bluer-when-brighter (BWB) trend.} 
    \label{fig:color-mag-type}
\end{figure*}

\subsection{Color-magnitude variation}
YSOs exhibit diverse types of color variation across the wavelength, e.g., ``bluer when brighter (BWB)", ``redder when brighter (RWB)," and cyclic trends, etc. \citet{2019AJ....158..240H} found Gaia 19ajj to show a BWB trend in the mid-infrared color variation, concluding that the blueing behavior was not only due to the clearing of the extinction but also intrinsic to the source. \citet{park21} studied the mid-infrared color variation of a large sample of YSOs and found evidence of YSOs showing monotonic color variations with RWB and BWB trends as well as YSOs with clockwise and anti-clockwise color variation. \citet{2024MNRAS.52711651A} studied a low luminosity protostar, SSTgbsJ214706, and presented that the observed color variation of the source is not uni-directional. Initially, it becomes redder, and as the brightness increases, the source turns bluer, showing the BWB trend. The source again becomes redder, showing a periodic variation in the color-magnitude diagram. These different trends of color variation with light curves correspond to various driving mechanisms of variables (see section \ref{sec:discussion}).

We have investigated (W1$–$W2) mid-infrared color variations of our sample for similar clear patterns. We have performed a linear least square fit of the color (W1$–$W2) and W2 mag diagram to estimate the slope using the Python Orthogonal distance regression (ODR) module \citep{boggs1990orthogonal}, considering the uncertainties on both axes. We also calculated the Pearson rank correlation coefficient and $p$-value for the null hypothesis. We then considered the source to show redder-when-brighter (RWB) if the slope is negative, while bluer-when-brighter (BWB) for a positive slope if the slope is measured better than 1$\sigma$ uncertainty (i.e., error in slope) and the $p$-value is less than 0.01. The number of RWB sources (4840 sources) is more common than BWB (826 sources) i.e., the BWB and RWB sources consist of 14.6$\pm$0.5\% and 85.4$\pm$0.5\% respectively, of the total sample. Most sources with a slope close to zero and a high $p$-value are expected to have no color variation. However, some may show complex or cycle variations in color. 

Figure \ref{fig:example_color} shows an example of BWB and RWB color variations. Regarding YSO classes, we find that the fraction of BWB sources to the total (BWB+RWB) decreases from 29.0$\pm$1.1\% for Class I to 9.8$\pm$0.5\% for Class II to 4.0$\pm$0.9\% for Class III. In Figure \ref{fig:color-mag-type}, we plot the color slope with the spectral index or evolutionary stage. We found that young sources exhibit more BWB trends than RWB, mainly due to their active accretion processes and greater variability during their early evolutionary stages. {\it Drop} sources tend to have more BWB (28.6$\pm$6.0\%) than {\it Burst} (7.6$\pm$2.3\%) sources. The differences arise from the nature of variability in each type, with {\it Drop} sources demonstrating systematic changes. In contrast, {\it Burst} sources exhibit more erratic behavior due to their episodic nature. {\it Drop} sources exhibit a BWB trend, likely due to a decrease in obscuration along the line of sight, revealing hotter inner regions of the system \citep{2013MNRAS.430.2910S, 2014ApJS..211....3P, 2015AJ....150..132R}. However, during {\it Burst} events, increased accretion leads to enhanced disk heating, resulting in stronger infrared re-emission. This reprocessed radiation can dominate the observed flux, causing the system to appear RWB at the peak of the outburst \citep{2013AJ....145...59H, 2014prpl.conf..387A, 2014Cody, apjad14f8bib10}.
On average, {\it Secular} variables (22.1$\pm$1.9\%) show fewer BWB trends than {\it Irregular} (29.3$\pm$1.1\%) sources. These statistics are consistent with \citet{10.1093/mnras/stae1601}.

\begin{table*}
    \centering
    \caption{Table containing variability statistics of individual objects.\\
    $^a$ The values are taken from \citet{2021ApJS..254...33K}. \\
    (This table is available in its entirety in FITS format.)}
    \begin{tabular}{l l l} \\ \hline \hline 
    Column & Column ID & Description \\ \hline
1 & Objname	        & Name of the YSOs$^a$    \\
2 & RA	            & ICRS R.A. coordinate in decimal degrees$^a$ \\
3 & DEC	            & ICRS decl. coordinate in decimal degrees$^a$ \\
4 & SED\_SLOPE	    & Spectral index for YSO class$^a$ \\
5 & YSO\_CLASS      & YSO class$^a$ \\
6 & No\_of\_point	& No of data points in WISE W2 band light curve  \\
7 & mean\_mag\_w2 	& mean magnitude in WISE W2 band light curve \\
8 & median\_mag\_w2 & median magnitude in WISE W2 band light curve	\\ 	 
9 & sig\_flux\_w2	& standard deviation (flux) in WISE W2 band light curve \\
10 & err\_flux\_w2	& uncertainty (flux) in WISE W2 band light curve \\
11 & del\_mag\_w2	& magnitude difference between brightest and faintest mag \\
12 & Period	        & LSP period \\
13 & FLP\_LSP\_BOOT	& Flase alarm probability in LSP bootstrapping \\ 
14 & slope	        & Slope in linear  fit to the light curve \\
15 & slope\_err	 	& Uncertainty in col. 14 \\
16 & r\_value	    & Spearman rank correction coefficient of col. 14 \\
17 & r\_value\_err	& Uncertainty in col. 16 \\
18 & p\_value	    & $p$-value of col. 14 \\
19 & p\_value\_err	& Uncertainty of col. 18  \\
20 & slope\_color	& Slope in color (W1-W2) vs magnitude W2 diagram \\
21 & eslope\_color	& Uncertainty of col 20. \\ 	 
22 & pearson\_coeff\_color & person rank correlation coefficient of col. 20	 \\
23 & p\_value\_color	 & $p$-value of col. 20	\\
24 & LC\_type            & Type of Light curve \\ \hline
    \end{tabular}
    \label{tab:var_stat_full_table}
\end{table*}

\section{Discussions}\label{sec:discussion}

\subsection{Reliability of variable YSO selection}
In this section, we evaluate the reliability of using the criteria $\Delta W2/\sigma >3$ and SD/$\sigma>3$ for identifying variable objects. We note that the number of variables is strongly affected by $\sigma$, which we calculated based on the equation \ref{eq:sig}. \citet{2023ApJ...958..135S} use a similar equation in their study and demonstrated, using a sample of non-variable active galactic nuclei (AGN), that this error should be reduced by a factor of $\sqrt{N_s}$, where $N_s$ represents the number of epochs in the short-term (six-month) light curve (see Section 3.3.3). Given that WISE has an orbital period of approximately 90 minutes and continuously observes the same region of the sky for about a day, our error estimates are, on average, overestimated by a factor of $\sim$3. To quantitatively assess this overestimation, we calculated the standard deviation of the light curves for a subsample of 1000 non-variable stars with  $\Delta W2 <$ 0.1 mag. The resulting average standard deviation is 0.02 mag, while the mean error, computed using equation \ref{eq:sig}, is 0.05 mag. This suggests that our errors are overestimated by a factor of 2–3.

To further assess the false positive rate, we performed a Monte Carlo simulation. For each selected object, we generated 10,000 artificial light curves by adding Gaussian noise with zero mean and a standard deviation equal to the original light curve’s uncertainty ($\sigma$, as defined in Section \ref{sec:data}) to its mean magnitude. The cadence and epochs were kept identical to the original light curve. We then evaluated how many of these synthetic light curves met our variability selection criteria: $\Delta W2/\sigma >3$ and SD$/\sigma>3$. This process was repeated for a randomly chosen subset of 1,000 objects from our final list of 20,893 YSOs (see section \ref{sec:data}). Our results show that, on average, 82\% of sources satisfied the condition, whereas no light curve met the condition. This confirms that the criterion SD/$\sigma>3$ effectively selects variable objects. Note that our final sample of variable YSOs consists of 5467 objects, among which 4803 meet both criteria, while the remaining sources were identified based on additional conditions outlined in Section 3.1, including the $\Delta W2/\sigma >3$ criterion.

\subsection{Comparison with previous studies}
The YSOVAR program provided dedicated Spitzer observations at 3.6 and 4.5 $\mu$m over timescales of $\sim$40 days for YSOs in the Orion cluster and 11 other star-forming regions \citep{2014AJ....148...92R}. \citet{apjad14f8bib22} used YSOVAR data to study the Orion cluster and found that 70\% of YSOs with disks and 44\% of Weak T-Tauri stars (WTTs) exhibited variability. Similarly, \citet{2018Wolk} investigated mid-infrared variability of YSOs in Serpens South, reporting variability fractions of 77\% $\pm$ 13\%, 74\% $\pm$ 20\%, and 59\% $\pm$ 19\% for SED classes I, flat, and II, respectively.

Mid-infrared variability studies with Spitzer and WISE have also revealed large-scale changes in protostellar luminosities over multi-year baselines. \citet{2013MNRAS.430.2910S} compared two epochs of Spitzer and WISE data for about 8000 YSOs and detected outbursts with amplitudes exceeding 1 mag over $\sim$5 years. Similarly, \citet{2019ApJ...872..183F} identified two outbursts in 319 protostars in the Orion molecular clouds over 6.5 years, while \citet{2014ApJ...782...51A} focused on EXor-type outbursts. \citet{2019ApJ...883....6U} monitored 331 massive protostars using NEOWISE and found five mid-infrared variable candidates, and \citet{2020MNRAS.499.1805L} analyzed WISE/NEOWISE data from 2010–2017 to identify high-amplitude variability in infrared-dark clouds, including a YSO outburst.

Our study extends these investigations by examining a significantly larger sample of YSOs over a longer timescale (8–10 years), encompassing both ALLWISE and NEOWISE epochs. We confirm the well-established trend that younger YSOs are more likely to exhibit variability, finding variability fractions of $36.3\pm0.6$\% for Class I, $22.1\pm0.4$\% for Class II, and $22.5\pm1$\% for Class III sources. Compared to YSOVAR, our variability fractions are lower, likely due to the long-term nature of our study and our six-month cadence light curves, which differ from YSOVAR's higher-cadence observations. Moreover, as mentioned above, the SPICY catalog \citep{2021ApJS..254...33K} that we have used was constructed from IRAC, which has higher sensitivity and spatial resolution compared to WISE.

Recent work by \citet{park21} analyzed 6.5 years of NEOWISE-R data for 5,398 YSOs and reported variability fractions of $\sim$55\% (Class 0/I and flat), $\sim$33\% (Class II), and $\sim$15\% (Class III), values that are intermediate between YSOVAR and our study. Similarly, \citet{2024Lee} used NEOWISE and YSOVAR data to study mid-infrared variability, reporting a higher variability detection rate in the YSOVAR short-term light curve that spans approximately a month (77.6\%) compared to long-term NEOWISE data that span several years (43.0\%), likely due to the higher sensitivity of YSOVAR. They also found that short-term variations are predominantly {\it Secular}, whereas long-term variations observed with NEOWISE are mostly stochastic. While both \citet{2024Lee} and our study used NEOWISE data, our sample size is larger, and our monitoring period extends beyond 10 years, providing a more comprehensive view of long-term mid-infrared variability. A direct comparison with \citet{2024Lee} was limited, as we found only one common YSO between our datasets (originally from \citealt{park21}), which had a similar $\Delta W2$ in both catalogs. The minimal overlap prevents detailed cross-validation.

\subsection{The variability mechanisms}
YSOs exhibit mid-infrared (3–5 $\micron$) variability due to dynamic interactions between dust and gas in the inner disk, driven by accretion, disk instabilities, and structural changes. Unlike optical and near-infrared variability, which primarily trace the stellar surface and magnetospheric processes, mid-infrared variations originate from circumstellar dust reprocessing stellar radiation. Long-term variability is often linked to disk geometry changes, accretion fluctuations, or structural evolution \citep{2014Cody, 2018Wolk, 2021MNRAS.504..830G}. Periodic variations may arise from inner disk clumps, warps, or binary interactions causing disk eclipses and gravitational perturbations \citep{2019MNRAS.482.5459M, apjac1745bib69}. Increasing mid-infrared flux may indicate heating due to enhanced accretion, while dimming suggests disk clearing or long-term obscuration \citep{2018Davies, 2021Covey}. Short-term variability is more stochastic, likely driven by episodic accretion bursts, disk turbulence, or transient obscuration by evolving dust structures \citep{apjac1745bib143, apjac1745bib126}. Unlike optical brightening linked to stellar magnetospheric changes, mid-infrared variations reflect thermal emission shifts in the disk, with dimming better explained by inner disk extinction rather than magnetospheric occultations \citep{2013Bouvier, apjac1745bib63}.

The mid-infrared variability detected in NEOWISE observations is limited to timescales of six months or longer, meaning short-period (days to weeks) fluctuations commonly observed in the optical/NIR are not well sampled \citep{2024Lee}. The linear variability trends in the mid-infrared could be indicative of long-duration outbursts akin to FUor events, where accretion-driven disk heating results in a sustained increase in infrared emission. Conversely, declining mid-infrared flux may correspond to gradual accretion decay, disk dissipation, or large-scale obscuration by circumstellar material \citep{park21}. Overall, our results emphasize the importance of disk processes in shaping mid-infrared variability. Detailed investigations of highly variable YSOs cataloged in this work, along with infrared spectroscopic observation, would help better infer the physical mechanisms at play within the inner regions of circumstellar disks, particularly in the context of YSO evolution and disk dynamics.

\section{Conclusion}\label{sec: conclusion}
We performed an infrared color and variability study of more than 20,000 candidate YSOs, selected from the SPICY catalog, using archival data from the WISE survey consisting of data from both the ALLWISE and NESOWISE spanning about a decade with a typical 6-month cadence. Our main conclusions are as follows:
\begin{enumerate}
\item Using the W2-band lightcurve, a total of 5467 YSOs, i.e., around one-fourth of the total sample, are found to be variable, with significant variation in the fraction by evolutionary classes, i.e., 36.3$\pm$0.6\% for Class I, 22.1$\pm$0.4\% for Class II, and 22.5$\pm$1\% for Class III YSOs, suggesting high variability fraction in younger sources. We found 930 sources with $\Delta W2 >$ 1 mag and 68 sources with $\Delta W2 >$2 mag.

\item We categorized the variables into six groups based on their variability patterns: three are classified as {\it Secular} ({\it Linear}, {\it Curved}, and {\it Periodic}), and three as {\it Stochastic} ({\it Burst}, {\it Drop}, and {\it Irregular}). {\it Irregular} variables dominate all of them with 19.7$\pm$0.3\% of the total sample. {\it Burst} YSOs are found to be relatively more common in the younger class (1.3$\pm$0.1\%) than {\it Drop} sources (0.5$\pm$0.1\%). On average, {\it Bursts} (0.65 mag) and {\it Drops} (0.54 mag) exhibit larger variability amplitudes (median) compared to the {\it Secular} class (0.42 mag).

\item Based on color-magnitude variability, YSOs show `BWB' and `RWB' trends; however, the RWB trend (85.4$\pm$0.5\%) is more common than BWB (14.6$\pm$0.5\%). Separating the sources into different classes, we found the fraction of BWB to the total (BWB+RWB) decreases as the sources become older, i.e., from 29.0$\pm$1.1\% to 4.0$\pm$0.9\% from Class I to Class III. These trends could be due to changes in accretion rates or extinction caused by obscuration.
\end{enumerate}

In this work, we present the complete catalog as a publicly available resource to support diverse research on the accretion processes and evolutionary pathways of YSOs, for example, through infrared spectroscopy and multi-band follow-up observations of highly variable YSOs.

\begin{acknowledgments}
The authors thank the reviewer for his/her valuable suggestions, which significantly improved the quality of the manuscript. This publication makes use of data products from the Wide-field Infrared Survey Explorer (WISE), which is a joint project of the University of California, Los Angeles, and the Jet Propulsion Laboratory/California Institute of Technology, and NEOWISE, which is a project of the Jet Propulsion Laboratory/California Institute of Technology. WISE and NEOWISE are funded by the National Aeronautics and Space Administration. This research has used the NASA/IPAC Infrared Science Archive, which is funded by the National Aeronautics and Space Administration and operated by the California Institute of Technology. NS acknowledges the financial support provided under the National Post-doctoral Fellowship (NPDF; File Number: PDF/2022/001040) by the Science \& Engineering Research Board (SERB), a statutory body of the Department of Science \& Technology (DST), Government of India. NS thanks Suvendu Rakshit for the useful discussions.
\end{acknowledgments}

\vspace{5mm}
\facilities{WISE, All-WISE, NEO-WISE}

\software{Astropy \citep{2013A&A...558A..33A, 2018AJ....156..123A,2022ApJ...935..167A}, Matplotlib \citep{2007CSE.....9...90H}, NumPy \citep{2020Natur.585..357H}, SciPy \citep{2020NatMe..17..261V}.    }

\bibliography{yso-var}{}

\begin{thebibliography}{}
\expandafter\ifx\csname natexlab\endcsname\relax\def\natexlab#1{#1}\fi
\providecommand{\url}[1]{\href{#1}{#1}}
\providecommand{\dodoi}[1]{doi:~\href{http://doi.org/#1}{\nolinkurl{#1}}}
\providecommand{\doeprint}[1]{\href{http://ascl.net/#1}{\nolinkurl{http://ascl.net/#1}}}
\providecommand{\doarXiv}[1]{\href{https://arxiv.org/abs/#1}{\nolinkurl{https://arxiv.org/abs/#1}}}

\bibitem[{{Antoniucci} {et~al.}(2014){Antoniucci}, {Giannini}, {Li Causi}, \& {Lorenzetti}}]{2014ApJ...782...51A}
{Antoniucci}, S., {Giannini}, T., {Li Causi}, G., \& {Lorenzetti}, D. 2014, \apj, 782, 51, \dodoi{10.1088/0004-637X/782/1/51}

\bibitem[{{Ashraf} {et~al.}(2024){Ashraf}, {Jose}, {Lee}, {Contreras Pe{\~n}a}, {Herczeg}, {Liu}, {Johnstone}, \& {Lee}}]{2024MNRAS.52711651A}
{Ashraf}, M., {Jose}, J., {Lee}, H.-G., {et~al.} 2024, \mnras, 527, 11651, \dodoi{10.1093/mnras/stad3900}

\bibitem[{{Astropy Collaboration} {et~al.}(2013){Astropy Collaboration}, {Robitaille}, {Tollerud}, {Greenfield}, {Droettboom}, {Bray}, {Aldcroft}, {Davis}, {Ginsburg}, {Price-Whelan}, {Kerzendorf}, {Conley}, {Crighton}, {Barbary}, {Muna}, {Ferguson}, {Grollier}, {Parikh}, {Nair}, {Unther}, {Deil}, {Woillez}, {Conseil}, {Kramer}, {Turner}, {Singer}, {Fox}, {Weaver}, {Zabalza}, {Edwards}, {Azalee Bostroem}, {Burke}, {Casey}, {Crawford}, {Dencheva}, {Ely}, {Jenness}, {Labrie}, {Lim}, {Pierfederici}, {Pontzen}, {Ptak}, {Refsdal}, {Servillat}, \& {Streicher}}]{2013A&A...558A..33A}
{Astropy Collaboration}, {Robitaille}, T.~P., {Tollerud}, E.~J., {et~al.} 2013, \aap, 558, A33, \dodoi{10.1051/0004-6361/201322068}

\bibitem[{{Astropy Collaboration} {et~al.}(2018){Astropy Collaboration}, {Price-Whelan}, {Sip{\H{o}}cz}, {G{\"u}nther}, {Lim}, {Crawford}, {Conseil}, {Shupe}, {Craig}, {Dencheva}, {Ginsburg}, {VanderPlas}, {Bradley}, {P{\'e}rez-Su{\'a}rez}, {de Val-Borro}, {Aldcroft}, {Cruz}, {Robitaille}, {Tollerud}, {Ardelean}, {Babej}, {Bach}, {Bachetti}, {Bakanov}, {Bamford}, {Barentsen}, {Barmby}, {Baumbach}, {Berry}, {Biscani}, {Boquien}, {Bostroem}, {Bouma}, {Brammer}, {Bray}, {Breytenbach}, {Buddelmeijer}, {Burke}, {Calderone}, {Cano Rodr{\'\i}guez}, {Cara}, {Cardoso}, {Cheedella}, {Copin}, {Corrales}, {Crichton}, {D'Avella}, {Deil}, {Depagne}, {Dietrich}, {Donath}, {Droettboom}, {Earl}, {Erben}, {Fabbro}, {Ferreira}, {Finethy}, {Fox}, {Garrison}, {Gibbons}, {Goldstein}, {Gommers}, {Greco}, {Greenfield}, {Groener}, {Grollier}, {Hagen}, {Hirst}, {Homeier}, {Horton}, {Hosseinzadeh}, {Hu}, {Hunkeler}, {Ivezi{\'c}}, {Jain}, {Jenness}, {Kanarek}, {Kendrew}, {Kern}, {Kerzendorf}, {Khvalko}, {King}, {Kirkby}, {Kulkarni},
  {Kumar}, {Lee}, {Lenz}, {Littlefair}, {Ma}, {Macleod}, {Mastropietro}, {McCully}, {Montagnac}, {Morris}, {Mueller}, {Mumford}, {Muna}, {Murphy}, {Nelson}, {Nguyen}, {Ninan}, {N{\"o}the}, {Ogaz}, {Oh}, {Parejko}, {Parley}, {Pascual}, {Patil}, {Patil}, {Plunkett}, {Prochaska}, {Rastogi}, {Reddy Janga}, {Sabater}, {Sakurikar}, {Seifert}, {Sherbert}, {Sherwood-Taylor}, {Shih}, {Sick}, {Silbiger}, {Singanamalla}, {Singer}, {Sladen}, {Sooley}, {Sornarajah}, {Streicher}, {Teuben}, {Thomas}, {Tremblay}, {Turner}, {Terr{\'o}n}, {van Kerkwijk}, {de la Vega}, {Watkins}, {Weaver}, {Whitmore}, {Woillez}, {Zabalza}, \& {Astropy Contributors}}]{2018AJ....156..123A}
{Astropy Collaboration}, {Price-Whelan}, A.~M., {Sip{\H{o}}cz}, B.~M., {et~al.} 2018, \aj, 156, 123, \dodoi{10.3847/1538-3881/aabc4f}

\bibitem[{{Astropy Collaboration} {et~al.}(2022){Astropy Collaboration}, {Price-Whelan}, {Lim}, {Earl}, {Starkman}, {Bradley}, {Shupe}, {Patil}, {Corrales}, {Brasseur}, {N{\"o}the}, {Donath}, {Tollerud}, {Morris}, {Ginsburg}, {Vaher}, {Weaver}, {Tocknell}, {Jamieson}, {van Kerkwijk}, {Robitaille}, {Merry}, {Bachetti}, {G{\"u}nther}, {Aldcroft}, {Alvarado-Montes}, {Archibald}, {B{\'o}di}, {Bapat}, {Barentsen}, {Baz{\'a}n}, {Biswas}, {Boquien}, {Burke}, {Cara}, {Cara}, {Conroy}, {Conseil}, {Craig}, {Cross}, {Cruz}, {D'Eugenio}, {Dencheva}, {Devillepoix}, {Dietrich}, {Eigenbrot}, {Erben}, {Ferreira}, {Foreman-Mackey}, {Fox}, {Freij}, {Garg}, {Geda}, {Glattly}, {Gondhalekar}, {Gordon}, {Grant}, {Greenfield}, {Groener}, {Guest}, {Gurovich}, {Handberg}, {Hart}, {Hatfield-Dodds}, {Homeier}, {Hosseinzadeh}, {Jenness}, {Jones}, {Joseph}, {Kalmbach}, {Karamehmetoglu}, {Ka{\l}uszy{\'n}ski}, {Kelley}, {Kern}, {Kerzendorf}, {Koch}, {Kulumani}, {Lee}, {Ly}, {Ma}, {MacBride}, {Maljaars}, {Muna}, {Murphy}, {Norman},
  {O'Steen}, {Oman}, {Pacifici}, {Pascual}, {Pascual-Granado}, {Patil}, {Perren}, {Pickering}, {Rastogi}, {Roulston}, {Ryan}, {Rykoff}, {Sabater}, {Sakurikar}, {Salgado}, {Sanghi}, {Saunders}, {Savchenko}, {Schwardt}, {Seifert-Eckert}, {Shih}, {Jain}, {Shukla}, {Sick}, {Simpson}, {Singanamalla}, {Singer}, {Singhal}, {Sinha}, {Sip{\H{o}}cz}, {Spitler}, {Stansby}, {Streicher}, {{\v{S}}umak}, {Swinbank}, {Taranu}, {Tewary}, {Tremblay}, {de Val-Borro}, {Van Kooten}, {Vasovi{\'c}}, {Verma}, {de Miranda Cardoso}, {Williams}, {Wilson}, {Winkel}, {Wood-Vasey}, {Xue}, {Yoachim}, {Zhang}, {Zonca}, \& {Astropy Project Contributors}}]{2022ApJ...935..167A}
{Astropy Collaboration}, {Price-Whelan}, A.~M., {Lim}, P.~L., {et~al.} 2022, \apj, 935, 167, \dodoi{10.3847/1538-4357/ac7c74}

\bibitem[{{Attridge} \& {Herbst}(1992)}]{1992Attridge}
{Attridge}, J.~M., \& {Herbst}, W. 1992, \apjl, 398, L61, \dodoi{10.1086/186577}

\bibitem[{{Audard} {et~al.}(2014){Audard}, {{\'A}brah{\'a}m}, {Dunham}, {Green}, {Grosso}, {Hamaguchi}, {Kastner}, {K{\'o}sp{\'a}l}, {Lodato}, {Romanova}, {Skinner}, {Vorobyov}, \& {Zhu}}]{2014prpl.conf..387A}
{Audard}, M., {{\'A}brah{\'a}m}, P., {Dunham}, M.~M., {et~al.} 2014, in Protostars and Planets VI, ed. H.~{Beuther}, R.~S. {Klessen}, C.~P. {Dullemond}, \& T.~{Henning}, 387--410, \dodoi{10.2458/azu_uapress_9780816531240-ch017}

\bibitem[{{Baek} {et~al.}(2020){Baek}, {MacFarlane}, {Lee}, {Stamatellos}, {Herczeg}, {Johnstone}, {Pe{\~n}a}, {Varricatt}, {Hodapp}, {Chen}, \& {Kang}}]{2020ApJ...895...27B}
{Baek}, G., {MacFarlane}, B.~A., {Lee}, J.-E., {et~al.} 2020, \apj, 895, 27, \dodoi{10.3847/1538-4357/ab8ad4}

\bibitem[{{Bino} {et~al.}(2023){Bino}, {Basu}, {Dey}, {Auddy}, {Muller}, \& {Vorobyov}}]{2023Bino}
{Bino}, G., {Basu}, S., {Dey}, R., {et~al.} 2023, arXiv e-prints, arXiv:2302.03742, \dodoi{10.48550/arXiv.2302.03742}

\bibitem[{Boggs \& Rogers(1990)}]{boggs1990orthogonal}
Boggs, P.~T., \& Rogers, J.~E. 1990, in Contemporary Mathematics, Vol. 112, Statistical Analysis of Measurement Error Models and Applications: Proceedings of the AMS-IMS-SIAM Joint Summer Research Conference Held June 10-16, 1989 (Providence, RI: American Mathematical Society), 186

\bibitem[{{Bouvier} {et~al.}(2013){Bouvier}, {Grankin}, {Ellerbroek}, {Bouy}, \& {Barrado}}]{2013Bouvier}
{Bouvier}, J., {Grankin}, K., {Ellerbroek}, L.~E., {Bouy}, H., \& {Barrado}, D. 2013, \aap, 557, A77, \dodoi{10.1051/0004-6361/201321389}

\bibitem[{{Cargile} {et~al.}(2008){Cargile}, {Stassun}, \& {Mathieu}}]{2008Cargile}
{Cargile}, P.~A., {Stassun}, K.~G., \& {Mathieu}, R.~D. 2008, \apj, 674, 329, \dodoi{10.1086/524346}

\bibitem[{{Carpenter} {et~al.}(2001){Carpenter}, {Hillenbrand}, \& {Skrutskie}}]{2001Carpenter}
{Carpenter}, J.~M., {Hillenbrand}, L.~A., \& {Skrutskie}, M.~F. 2001, \aj, 121, 3160, \dodoi{10.1086/321086}

\bibitem[{{Choi} \& {Herbst}(1996)}]{1996Choi}
{Choi}, P.~I., \& {Herbst}, W. 1996, \aj, 111, 283, \dodoi{10.1086/117780}

\bibitem[{{Cody} {et~al.}(2014){Cody}, {Stauffer}, {Baglin}, {Micela}, {Rebull}, {Flaccomio}, {Morales-Calder{\'o}n}, {Aigrain}, {Bouvier}, {Hillenbrand}, {Gutermuth}, {Song}, {Turner}, {Alencar}, {Zwintz}, {Plavchan}, {Carpenter}, {Findeisen}, {Carey}, {Terebey}, {Hartmann}, {Calvet}, {Teixeira}, {Vrba}, {Wolk}, {Covey}, {Poppenhaeger}, {G{\"u}nther}, {Forbrich}, {Whitney}, {Affer}, {Herbst}, {Hora}, {Barrado}, {Holtzman}, {Marchis}, {Wood}, {Medeiros Guimar{\~a}es}, {Lillo Box}, {Gillen}, {McQuillan}, {Espaillat}, {Allen}, {D'Alessio}, \& {Favata}}]{2014Cody}
{Cody}, A.~M., {Stauffer}, J., {Baglin}, A., {et~al.} 2014, \aj, 147, 82, \dodoi{10.1088/0004-6256/147/4/82}

\bibitem[{{Contreras Pe{\~n}a} {et~al.}(2020){Contreras Pe{\~n}a}, {Johnstone}, {Baek}, {Herczeg}, {Mairs}, {Scholz}, {Lee}, \& {JCMT Transient Team}}]{Contreras20}
{Contreras Pe{\~n}a}, C., {Johnstone}, D., {Baek}, G., {et~al.} 2020, \mnras, 495, 3614, \dodoi{10.1093/mnras/staa1254}

\bibitem[{{Covey} {et~al.}(2021){Covey}, {Larson}, {Herczeg}, \& {Manara}}]{2021Covey}
{Covey}, K.~R., {Larson}, K.~A., {Herczeg}, G.~J., \& {Manara}, C.~F. 2021, \aj, 161, 61, \dodoi{10.3847/1538-3881/abcc73}

\bibitem[{{Davies} {et~al.}(2018){Davies}, {Kreplin}, {Kluska}, {Hone}, \& {Kraus}}]{2018Davies}
{Davies}, C.~L., {Kreplin}, A., {Kluska}, J., {Hone}, E., \& {Kraus}, S. 2018, \mnras, 474, 5406, \dodoi{10.1093/mnras/stx3150}

\bibitem[{Efron \& Tibshirani(1993)}]{efron1993bootstrap}
Efron, B., \& Tibshirani, R.~J. 1993, An Introduction to the Bootstrap (New York: Chapman \& Hall/CRC)

\bibitem[{{Fischer} {et~al.}(2019){Fischer}, {Safron}, \& {Megeath}}]{2019ApJ...872..183F}
{Fischer}, W.~J., {Safron}, E., \& {Megeath}, S.~T. 2019, \apj, 872, 183, \dodoi{10.3847/1538-4357/ab01dc}

\bibitem[{{Guo} {et~al.}(2021){Guo}, {Lucas}, {Contreras Pe{\~n}a}, {Smith}, {Morris}, {Kurtev}, {Borissova}, {Alonso-Garc{\'\i}a}, {Minniti}, {Chen{\'e}}, {Kumar}, {Caratti o Garatti}, {Froebrich}, \& {Stimson}}]{2021MNRAS.504..830G}
{Guo}, Z., {Lucas}, P.~W., {Contreras Pe{\~n}a}, C., {et~al.} 2021, \mnras, 504, 830, \dodoi{10.1093/mnras/stab882}

\bibitem[{{Harris} {et~al.}(2020){Harris}, {Millman}, {van der Walt}, {Gommers}, {Virtanen}, {Cournapeau}, {Wieser}, {Taylor}, {Berg}, {Smith}, {Kern}, {Picus}, {Hoyer}, {van Kerkwijk}, {Brett}, {Haldane}, {del R{\'\i}o}, {Wiebe}, {Peterson}, {G{\'e}rard-Marchant}, {Sheppard}, {Reddy}, {Weckesser}, {Abbasi}, {Gohlke}, \& {Oliphant}}]{2020Natur.585..357H}
{Harris}, C.~R., {Millman}, K.~J., {van der Walt}, S.~J., {et~al.} 2020, \nat, 585, 357, \dodoi{10.1038/s41586-020-2649-2}

\bibitem[{{Herbst} {et~al.}(1994){Herbst}, {Herbst}, {Grossman}, \& {Weinstein}}]{1994Herbst}
{Herbst}, W., {Herbst}, D.~K., {Grossman}, E.~J., \& {Weinstein}, D. 1994, \aj, 108, 1906, \dodoi{10.1086/117204}

\bibitem[{Herbst {et~al.}(1994)Herbst, Herbst, Grossman, \& Weinstein}]{apjac1745bib63}
Herbst, W., Herbst, D.~K., Grossman, E.~J., \& Weinstein, D. 1994, AJ, 108, 1906, \dodoi{10.1086/117204}

\bibitem[{{Hillenbrand} {et~al.}(2019){Hillenbrand}, {Reipurth}, {Connelley}, {Cutri}, \& {Isaacson}}]{2019AJ....158..240H}
{Hillenbrand}, L.~A., {Reipurth}, B., {Connelley}, M., {Cutri}, R.~M., \& {Isaacson}, H. 2019, \aj, 158, 240, \dodoi{10.3847/1538-3881/ab4e16}

\bibitem[{{Hillenbrand} {et~al.}(2013){Hillenbrand}, {Miller}, {Covey}, {Carpenter}, {Cenko}, {Silverman}, {Muirhead}, {Fischer}, {Crepp}, {Bloom}, \& {Filippenko}}]{2013AJ....145...59H}
{Hillenbrand}, L.~A., {Miller}, A.~A., {Covey}, K.~R., {et~al.} 2013, \aj, 145, 59, \dodoi{10.1088/0004-6256/145/3/59}

\bibitem[{Hodapp {et~al.}(2012)Hodapp, Chini, Watermann, \& Lemke}]{apjac1745bib69}
Hodapp, K.~W., Chini, R., Watermann, R., \& Lemke, R. 2012, ApJ, 744, 56, \dodoi{10.1088/0004-637X/744/1/56}

\bibitem[{{Hunter}(2007)}]{2007CSE.....9...90H}
{Hunter}, J.~D. 2007, Computing in Science and Engineering, 9, 90, \dodoi{10.1109/MCSE.2007.55}

\bibitem[{{Hunter} {et~al.}(2018){Hunter}, {Brogan}, {MacLeod}, {Cyganowski}, {Chibueze}, {Friesen}, {Hirota}, {Smits}, {Chandler}, \& {Indebetouw}}]{2018ApJ...854..170H}
{Hunter}, T.~R., {Brogan}, C.~L., {MacLeod}, G.~C., {et~al.} 2018, \apj, 854, 170, \dodoi{10.3847/1538-4357/aaa962}

\bibitem[{{Johnstone} {et~al.}(2013){Johnstone}, {Hendricks}, {Herczeg}, \& {Bruderer}}]{2013ApJ...765..133J}
{Johnstone}, D., {Hendricks}, B., {Herczeg}, G.~J., \& {Bruderer}, S. 2013, \apj, 765, 133, \dodoi{10.1088/0004-637X/765/2/133}

\bibitem[{{K{\'o}sp{\'a}l} {et~al.}(2007){K{\'o}sp{\'a}l}, {{\'A}brah{\'a}m}, {Prusti}, {Acosta-Pulido}, {Hony}, {Mo{\'o}r}, \& {Siebenmorgen}}]{2007A&A...470..211K}
{K{\'o}sp{\'a}l}, {\'A}., {{\'A}brah{\'a}m}, P., {Prusti}, T., {et~al.} 2007, \aap, 470, 211, \dodoi{10.1051/0004-6361:20066108}

\bibitem[{{Kuhn} {et~al.}(2021){Kuhn}, {de Souza}, {Krone-Martins}, {Castro-Ginard}, {Ishida}, {Povich}, {Hillenbrand}, \& {COIN Collaboration}}]{2021ApJS..254...33K}
{Kuhn}, M.~A., {de Souza}, R.~S., {Krone-Martins}, A., {et~al.} 2021, \apjs, 254, 33, \dodoi{10.3847/1538-4365/abe465}

\bibitem[{{Lee} {et~al.}(2024){Lee}, {Lee}, {Contreras Pe{\~n}a}, {Johnstone}, {Herczeg}, \& {Lee}}]{2024Lee}
{Lee}, S., {Lee}, J.-E., {Contreras Pe{\~n}a}, C., {et~al.} 2024, \apj, 962, 38, \dodoi{10.3847/1538-4357/ad14f8}

\bibitem[{{Lee} {et~al.}(2021){Lee}, {Johnstone}, {Lee}, {Herczeg}, {Mairs}, {Contreras-Pe{\~n}a}, {Hatchell}, {Naylor}, {Bell}, {Bourke}, {Broughton}, {Francis}, {Gupta}, {Harsono}, {Liu}, {Park}, {Plovie}, {Moriarty-Schieven}, {Scholz}, {Sharma}, {Teixeira}, {Wang}, {Aikawa}, {Bower}, {Vivien Chen}, {Bae}, {Baek}, {Chapman}, {Ping Chen}, {Du}, {Dutta}, {Forbrich}, {Guo}, {Inutsuka}, {Kang}, {Kirk}, {Kuan}, {Kwon}, {Lai}, {Lalchand}, {Lane}, {Lee}, {Liu}, {Morata}, {Pearson}, {Pon}, {Sahu}, {Shang}, {Stamatellos}, {Tang}, {Xu}, {Yoo}, \& {Rawlings}}]{2021LeeYH}
{Lee}, Y.-H., {Johnstone}, D., {Lee}, J.-E., {et~al.} 2021, \apj, 920, 119, \dodoi{10.3847/1538-4357/ac1679}

\bibitem[{Li \& Wang(2024)}]{10.1093/mnras/stae1601}
Li, J., \& Wang, T. 2024, Monthly Notices of the Royal Astronomical Society, 532, 2683, \dodoi{10.1093/mnras/stae1601}

\bibitem[{{Liu} {et~al.}(2018){Liu}, {Su}, {Zinchenko}, {Wang}, \& {Wang}}]{2018ApJ...863L..12L}
{Liu}, S.-Y., {Su}, Y.-N., {Zinchenko}, I., {Wang}, K.-S., \& {Wang}, Y. 2018, \apjl, 863, L12, \dodoi{10.3847/2041-8213/aad63a}

\bibitem[{{Lucas} {et~al.}(2020){Lucas}, {Elias}, {Points}, {Guo}, {Smith}, {Stecklum}, {Vorobyov}, {Morris}, {Borissova}, {Kurtev}, {Contreras Pe{\~n}a}, {Medina}, {Minniti}, {Ivanov}, \& {Saito}}]{2020MNRAS.499.1805L}
{Lucas}, P.~W., {Elias}, J., {Points}, S., {et~al.} 2020, \mnras, 499, 1805, \dodoi{10.1093/mnras/staa2915}

\bibitem[{{MacFarlane} {et~al.}(2019){MacFarlane}, {Stamatellos}, {Johnstone}, {Herczeg}, {Baek}, {Chen}, {Kang}, \& {Lee}}]{2019MNRAS.487.5106M}
{MacFarlane}, B., {Stamatellos}, D., {Johnstone}, D., {et~al.} 2019, \mnras, 487, 5106, \dodoi{10.1093/mnras/stz1512}

\bibitem[{{Mainzer} {et~al.}(2011){Mainzer}, {Bauer}, {Grav}, {Masiero}, {Cutri}, {Dailey}, {Eisenhardt}, {McMillan}, {Wright}, {Walker}, {Jedicke}, {Spahr}, {Tholen}, {Alles}, {Beck}, {Brandenburg}, {Conrow}, {Evans}, {Fowler}, {Jarrett}, {Marsh}, {Masci}, {McCallon}, {Wheelock}, {Wittman}, {Wyatt}, {DeBaun}, {Elliott}, {Elsbury}, {Gautier}, {Gomillion}, {Leisawitz}, {Maleszewski}, {Micheli}, \& {Wilkins}}]{2011ApJ...731...53M}
{Mainzer}, A., {Bauer}, J., {Grav}, T., {et~al.} 2011, \apj, 731, 53, \dodoi{10.1088/0004-637X/731/1/53}

\bibitem[{{Mainzer} {et~al.}(2014){Mainzer}, {Bauer}, {Cutri}, {Grav}, {Masiero}, {Beck}, {Clarkson}, {Conrow}, {Dailey}, {Eisenhardt}, {Fabinsky}, {Fajardo-Acosta}, {Fowler}, {Gelino}, {Grillmair}, {Heinrichsen}, {Kendall}, {Kirkpatrick}, {Liu}, {Masci}, {McCallon}, {Nugent}, {Papin}, {Rice}, {Royer}, {Ryan}, {Sevilla}, {Sonnett}, {Stevenson}, {Thompson}, {Wheelock}, {Wiemer}, {Wittman}, {Wright}, \& {Yan}}]{2014ApJ...792...30M}
{Mainzer}, A., {Bauer}, J., {Cutri}, R.~M., {et~al.} 2014, \apj, 792, 30, \dodoi{10.1088/0004-637X/792/1/30}

\bibitem[{{Meyer} {et~al.}(2019){Meyer}, {Vorobyov}, {Elbakyan}, {Stecklum}, {Eisl{\"o}ffel}, \& {Sobolev}}]{2019MNRAS.482.5459M}
{Meyer}, D.~M.~A., {Vorobyov}, E.~I., {Elbakyan}, V.~G., {et~al.} 2019, \mnras, 482, 5459, \dodoi{10.1093/mnras/sty2980}

\bibitem[{{Morales-Calder{\'o}n} {et~al.}(2011){Morales-Calder{\'o}n}, {Stauffer}, {Hillenbrand}, {Gutermuth}, {Song}, {Rebull}, {Plavchan}, {Carpenter}, {Whitney}, {Covey}, {Alves de Oliveira}, {Winston}, {McCaughrean}, {Bouvier}, {Guieu}, {Vrba}, {Holtzman}, {Marchis}, {Hora}, {Wasserman}, {Terebey}, {Megeath}, {Guinan}, {Forbrich}, {Hu{\'e}lamo}, {Riviere-Marichalar}, {Barrado}, {Stapelfeldt}, {Hern{\'a}ndez}, {Allen}, {Ardila}, {Bayo}, {Favata}, {James}, {Werner}, \& {Wood}}]{morales11}
{Morales-Calder{\'o}n}, M., {Stauffer}, J.~R., {Hillenbrand}, L.~A., {et~al.} 2011, \apj, 733, 50, \dodoi{10.1088/0004-637X/733/1/50}

\bibitem[{Morales-Calderón {et~al.}(2011)Morales-Calderón, Stauffer, Hillenbrand, {et~al.}}]{apjad14f8bib22}
Morales-Calderón, M., Stauffer, J.~R., Hillenbrand, L.~A., {et~al.} 2011, ApJ, 733, 50, \dodoi{10.1088/0004-637X/733/1/50}

\bibitem[{{Park} {et~al.}(2021){Park}, {Lee}, {Contreras Pe{\~n}a}, {Johnstone}, {Herczeg}, {Lee}, {Lee}, {Bhardwaj}, \& {Moriarty-Schieven}}]{park21}
{Park}, W., {Lee}, J.-E., {Contreras Pe{\~n}a}, C., {et~al.} 2021, \apj, 920, 132, \dodoi{10.3847/1538-4357/ac1745}

\bibitem[{{Parks} {et~al.}(2014){Parks}, {Plavchan}, {White}, \& {Gee}}]{2014ApJS..211....3P}
{Parks}, J.~R., {Plavchan}, P., {White}, R.~J., \& {Gee}, A.~H. 2014, \apjs, 211, 3, \dodoi{10.1088/0067-0049/211/1/3}

\bibitem[{Peña {et~al.}(2017)Peña, Lucas, Minniti, {et~al.}}]{apjad14f8bib10}
Peña, C.~C., Lucas, P.~W., Minniti, D., {et~al.} 2017, MNRAS, 465, 3011, \dodoi{10.1093/mnras/stw2801}

\bibitem[{{Rebull} {et~al.}(2014){Rebull}, {Cody}, {Covey}, {G{\"u}nther}, {Hillenbrand}, {Plavchan}, {Poppenhaeger}, {Stauffer}, {Wolk}, {Gutermuth}, {Morales-Calder{\'o}n}, {Song}, {Barrado}, {Bayo}, {James}, {Hora}, {Vrba}, {Alves de Oliveira}, {Bouvier}, {Carey}, {Carpenter}, {Favata}, {Flaherty}, {Forbrich}, {Hernandez}, {McCaughrean}, {Megeath}, {Micela}, {Smith}, {Terebey}, {Turner}, {Allen}, {Ardila}, {Bouy}, \& {Guieu}}]{2014AJ....148...92R}
{Rebull}, L.~M., {Cody}, A.~M., {Covey}, K.~R., {et~al.} 2014, \aj, 148, 92, \dodoi{10.1088/0004-6256/148/5/92}

\bibitem[{{Rice} {et~al.}(2015){Rice}, {Reipurth}, {Wolk}, {Vaz}, \& {Cross}}]{2015AJ....150..132R}
{Rice}, T.~S., {Reipurth}, B., {Wolk}, S.~J., {Vaz}, L.~P., \& {Cross}, N.~J.~G. 2015, \aj, 150, 132, \dodoi{10.1088/0004-6256/150/4/132}

\bibitem[{{Romanova} {et~al.}(2013){Romanova}, {Ustyugova}, {Koldoba}, \& {Lovelace}}]{2013Romanova}
{Romanova}, M.~M., {Ustyugova}, G.~V., {Koldoba}, A.~V., \& {Lovelace}, R.~V.~E. 2013, \mnras, 430, 699, \dodoi{10.1093/mnras/sts670}

\bibitem[{Rostopchina {et~al.}(2007)Rostopchina, Grinin, Shakhovskoi, Lomach, \& Minikulov}]{apjac1745bib126}
Rostopchina, A.~N., Grinin, V.~P., Shakhovskoi, D.~N., Lomach, A.~A., \& Minikulov, N.~K. 2007, ARep, 51, 55, \dodoi{10.1134/S1063772907010064}

\bibitem[{{Safron} {et~al.}(2015){Safron}, {Fischer}, {Megeath}, {Furlan}, {Stutz}, {Stanke}, {Billot}, {Rebull}, {Tobin}, {Ali}, {Allen}, {Booker}, {Watson}, \& {Wilson}}]{2015ApJ...800L...5S}
{Safron}, E.~J., {Fischer}, W.~J., {Megeath}, S.~T., {et~al.} 2015, \apjl, 800, L5, \dodoi{10.1088/2041-8205/800/1/L5}

\bibitem[{Scargle(1989)}]{apjad14f8bib27}
Scargle, J.~D. 1989, ApJ, 343, 874, \dodoi{10.1086/167757}

\bibitem[{{Scholz} {et~al.}(2013){Scholz}, {Froebrich}, \& {Wood}}]{2013MNRAS.430.2910S}
{Scholz}, A., {Froebrich}, D., \& {Wood}, K. 2013, \mnras, 430, 2910, \dodoi{10.1093/mnras/stt091}

\bibitem[{{Son} {et~al.}(2023){Son}, {Kim}, \& {Ho}}]{2023ApJ...958..135S}
{Son}, S., {Kim}, M., \& {Ho}, L.~C. 2023, \apj, 958, 135, \dodoi{10.3847/1538-4357/ad01bc}

\bibitem[{{Stassun} {et~al.}(2004){Stassun}, {Mathieu}, {Vaz}, {Stroud}, \& {Vrba}}]{2004Stassun}
{Stassun}, K.~G., {Mathieu}, R.~D., {Vaz}, L. P.~R., {Stroud}, N., \& {Vrba}, F.~J. 2004, \apjs, 151, 357, \dodoi{10.1086/382353}

\bibitem[{Stauffer {et~al.}(2014)Stauffer, Cody, Baglin, {et~al.}}]{apjad14f8bib31}
Stauffer, J., Cody, A.~M., Baglin, A., {et~al.} 2014, AJ, 147, 83, \dodoi{10.1088/0004-6256/147/4/83}

\bibitem[{{Stauffer} {et~al.}(2014){Stauffer}, {Cody}, {Baglin}, {Alencar}, {Rebull}, {Hillenbrand}, {Venuti}, {Turner}, {Carpenter}, {Plavchan}, {Findeisen}, {Carey}, {Terebey}, {Morales-Calder{\'o}n}, {Bouvier}, {Micela}, {Flaccomio}, {Song}, {Gutermuth}, {Hartmann}, {Calvet}, {Whitney}, {Barrado}, {Vrba}, {Covey}, {Herbst}, {Furesz}, {Aigrain}, \& {Favata}}]{2014Stauffer}
{Stauffer}, J., {Cody}, A.~M., {Baglin}, A., {et~al.} 2014, \aj, 147, 83, \dodoi{10.1088/0004-6256/147/4/83}

\bibitem[{Takasao {et~al.}(2019)Takasao, Tomida, Iwasaki, \& Suzuki}]{apjac1745bib143}
Takasao, S., Tomida, K., Iwasaki, K., \& Suzuki, T.~K. 2019, ApJL, 878, L10, \dodoi{10.3847/2041-8213/ab22bb}

\bibitem[{{Uchiyama} \& {Ichikawa}(2019)}]{2019ApJ...883....6U}
{Uchiyama}, M., \& {Ichikawa}, K. 2019, \apj, 883, 6, \dodoi{10.3847/1538-4357/ab372e}

\bibitem[{{Virtanen} {et~al.}(2020){Virtanen}, {Gommers}, {Oliphant}, {Haberland}, {Reddy}, {Cournapeau}, {Burovski}, {Peterson}, {Weckesser}, {Bright}, {van der Walt}, {Brett}, {Wilson}, {Millman}, {Mayorov}, {Nelson}, {Jones}, {Kern}, {Larson}, {Carey}, {Polat}, {Feng}, {Moore}, {VanderPlas}, {Laxalde}, {Perktold}, {Cimrman}, {Henriksen}, {Quintero}, {Harris}, {Archibald}, {Ribeiro}, {Pedregosa}, {van Mulbregt}, \& {SciPy 1. 0 Contributors}}]{2020NatMe..17..261V}
{Virtanen}, P., {Gommers}, R., {Oliphant}, T.~E., {et~al.} 2020, Nature Methods, 17, 261, \dodoi{10.1038/s41592-019-0686-2}

\bibitem[{Wilson(1927)}]{Wilson01061927}
Wilson, E.~B. 1927, Journal of the American Statistical Association, 22, 209, \dodoi{10.1080/01621459.1927.10502953}

\bibitem[{{Wolk} {et~al.}(2018){Wolk}, {G{\"u}nther}, {Poppenhaeger}, {Winston}, {Rebull}, {Stauffer}, {Gutermuth}, {Cody}, {Hillenbrand}, {Plavchan}, {Covey}, \& {Song}}]{2018Wolk}
{Wolk}, S.~J., {G{\"u}nther}, H.~M., {Poppenhaeger}, K., {et~al.} 2018, \aj, 155, 99, \dodoi{10.3847/1538-3881/aaa6c4}

\bibitem[{{Wright} {et~al.}(2010){Wright}, {Eisenhardt}, {Mainzer}, {Ressler}, {Cutri}, {Jarrett}, {Kirkpatrick}, {Padgett}, {McMillan}, {Skrutskie}, {Stanford}, {Cohen}, {Walker}, {Mather}, {Leisawitz}, {Gautier}, {McLean}, {Benford}, {Lonsdale}, {Blain}, {Mendez}, {Irace}, {Duval}, {Liu}, {Royer}, {Heinrichsen}, {Howard}, {Shannon}, {Kendall}, {Walsh}, {Larsen}, {Cardon}, {Schick}, {Schwalm}, {Abid}, {Fabinsky}, {Naes}, \& {Tsai}}]{2010AJ....140.1868W}
{Wright}, E.~L., {Eisenhardt}, P. R.~M., {Mainzer}, A.~K., {et~al.} 2010, \aj, 140, 1868, \dodoi{10.1088/0004-6256/140/6/1868}

\end{thebibliography}
\bibliographystyle{aasjournal}

\end{document}